\documentclass[a4paper,10pt,twoside]{cpc-hepnp}
\usepackage{multicol}
\usepackage{graphicx}
\usepackage{booktabs}
\usepackage{amssymb,bm,mathrsfs,bbm,amscd}
\usepackage[tbtags]{amsmath}
\usepackage{lastpage}

\begin{document}

\fancyhead[c]{\small Chinese Physics C~~~Vol. xx, No. x (201x) xxxxxx}
\fancyfoot[C]{\small 010201-\thepage}

\footnotetext[0]{Received xx xx 2016}

\title{Measurement Of Energy Resolution For TES Microcalorimeter With Optical Pulses\thanks{CAS Strategic Priority A(Grant No. XDA040762)}}

\author{%
      Shuo Zhang $^{3;1)}$
\quad Qing-Ya Zhang $^{1,2;1)}$%
\\
\quad Jianshe Liu $^{1,2}$%
\quad CongZhan Liu $^{3}$\email{liucz@ihep.ac.cn}
\quad Wenhui Dong $^{1,2}$%
\quad Wei Chen $^{1,2}$\email{weichen@mail.tsinghua.edu.cn}%
}
\maketitle

\address{%
$^1$ {Tsinghua National Laboratory for Information Science and Technology, Tsinghua University, Beijing 100084, China}

$^2$ {Institute of Microelectronics, Department of Micro/Nanoelectronics, Tsinghua University, Beijing 100084, China}

$^3$ {Key Laboratory of Particle Astrophysics, Institute of High Energy Physics, Chinese Academy Of Sciences, Beijing, 100049, China}
}

\begin{abstract}
Energy resolution is an important figure of merit for TES microcalorimeter. We propose a laser system to measure the energy resolution of TES microcalorimeter with a 1550 nm laser source. Compared to method that characterizes the performance by irradiating the detector using X-ray photons from a radioactive source placed inside the refrigerator, our system is safer and more convenient. The feasibility of this system has been demonstrated in the measurement of an Al/Ti bilayer TES microcalorimeter. In this experiment, the tested detector showed a energy resolution of 72 eV in the energy range from 0.2 keV to 0.9 keV.
\end{abstract}



\footnotetext[0]{\hspace*{-3mm}\raisebox{0.3ex}{$\scriptstyle\copyright$}2013
Chinese Physical Society and the Institute of High Energy Physics
of the Chinese Academy of Sciences and the Institute
of Modern Physics of the Chinese Academy of Sciences and IOP Publishing Ltd}%

\begin{multicols}{2}

\section{Introduction}

TES (Transition Edge Sensor) microcalorimeter, as a very sensitive temperature sensor, has been widely used in radiation detection. Compared to the traditional semi-conductor detectors, TES microcalorimeter has excellent detection efficiency and much higher energy resolution. TES Microcalorimeter has been playing an important role in the field of X-ray astronomy \cite{astronomyapply1,astronomyapply2,astronomyapply3,astronomyapply4}, nuclear energy spectrum\cite{astronomyapply5}, X-ray  spectroscopy analysis\cite{astronomyapply6}, elemental analysis\cite{astronomyapply6}, neutrino mass measurement\cite{astronomyapply7}, dark matter detection and so on.

Generally, the energy resolution of TES microcalorimeter is measured by radioactive sources whose energy are limited and discrete. If we need to measure the energy resolution at other energy, we should interrupt the operation of the refrigerator and change the radioactive sources. This process take a lot of time and will change the test environment.

Microcalorimeter detects X-ray base on the thermal signals, so we can use a thermal source replace radioactive source. We use a 1550 nm laser source to replace the radioactive source as the energy impulse for the test of TES. In comparing with the radioactive source, the equivalent energy and the frequency of laser pulse is adjustable. We can therefore use the laser to find the proper work point of the TES and get the energy resolution, linear dynamic range, characteristic shape of signal pulse very quickly. In this paper, we will introduce the system, and demonstrate the feasibility of this system by measuring the energy resolution of an experimental TES microcalorimeter we fabricated.

\subsection{Introduction to the laser test system}
As shown in Figure \ref{fig1}, a 1550 nm laser source with a constant output power $P_{0}$ is used to mimic the TES microcalorimeter, the pulse width $t$ and frequency $f$ modulated by a function generator. Current change signal $\delta{I}$ is converted to a voltage signal by SQUID amplifier, after filtered by pre-amplifier and Liner amplifier the signal is analysed by a Multi-channel analyser, then we got the energy spectrum of the laser pulse.

The bias voltage for the TES is provided by a room temperature current source $I_b$ and a 8.1 m$\Omega$ resistor ($R_b$) at 2.5 K. The TES branch has a parasitic resistance $R_p$ = 44m$\Omega$ from the wirings. The input inductance of SQUID amplifier ($L_{in}$) is about 600 nH.

\begin{center}
\includegraphics[width=8cm]{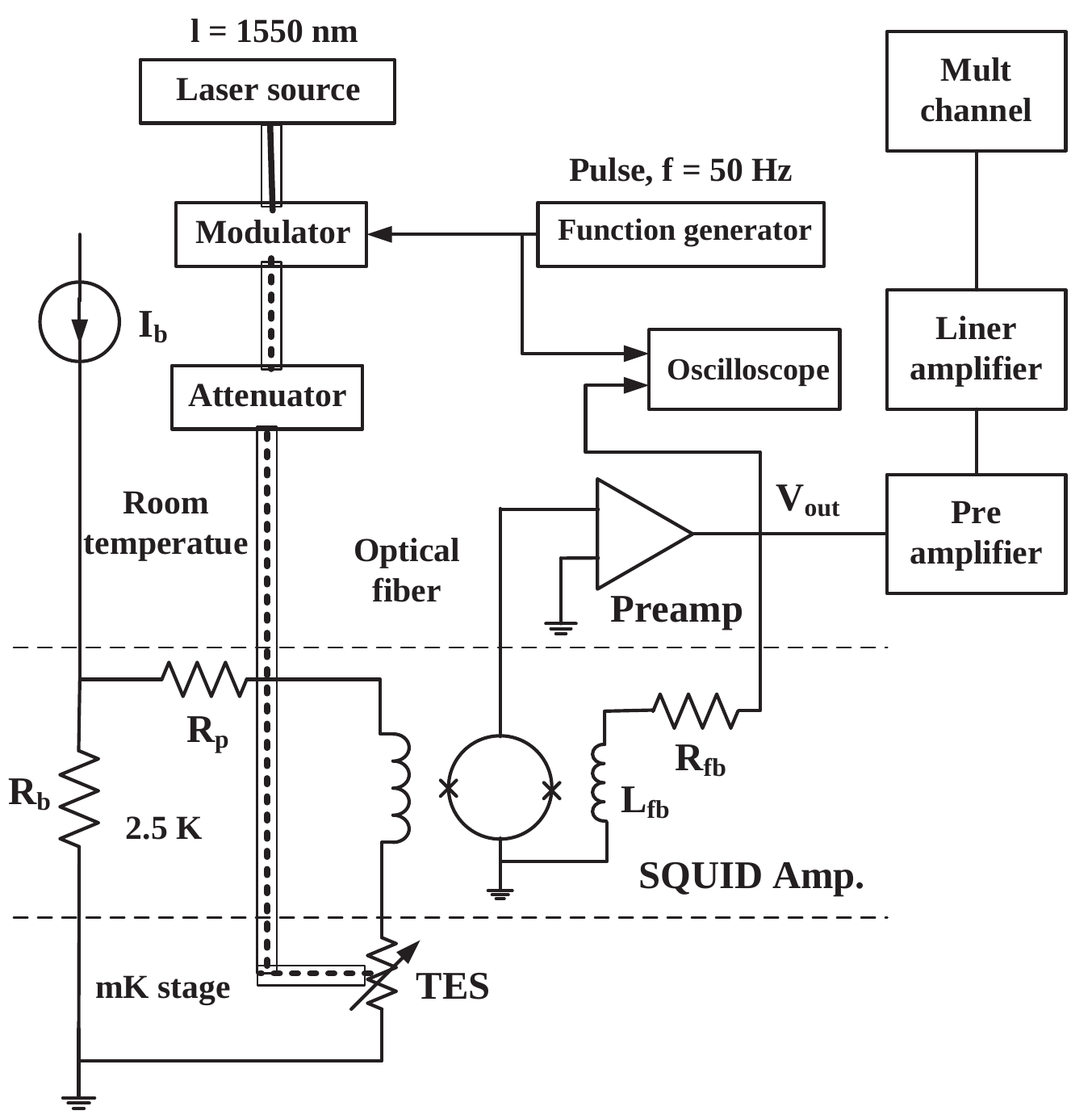}
\figcaption{\label{fig1}   The diagram of TES testing system using optical laser pulses.}
\end{center}

\subsection{analysis on energy spectrum}
As shown in Eq (\ref{eq1}), keeping $P_0$ constant£¬we can adjust photon number $n_{t}$ and average energy of laser pulse $E(\overline{n_{t}})$ by controlling $t$, here $E_{\lambda}$ is energy of single photon, for laser with wavelength of 1550nm, $E_{\lambda}=0.8eV$.
\begin{equation}\label{eq1}
  E(\overline{n_{t}})=\overline{n_{t}}*E_{\lambda}=k*P_{0}*t
\end{equation}

Compare to radiation sources, energy of laser pulse is not constant. Full width of half hight $FWHM_{\overline{n_t}}$ from Multi-channel analyser have two components£ºfluctuations caused by input photon number $\delta_{\lambda}^{2}$ and fluctuation caused by energy resolution of TES microcalorimeter itself $\delta_{r}^{2}$.
\begin{equation}\label{eq2}
FWHM_{n_t}=2.35*\delta_{n_t}=2.35*\sqrt{\overline{n_{t}}*\delta_{\lambda}^{2}+\delta_{r}^{2}}
\end{equation}

Although it is not constant, photon number in a laser pulse obey Poisson distribution $\pi(\overline{n_{t}})$, since the energy of laser photon is pretty small compare to X-ray source, $\overline{n_{t}}$ should be a very big number£¬so we can simplify it to a Gauss distribution $N(\overline{n_{t}},\overline{n_{t}}*\delta_{\lambda}^{2})$£¬where $\delta_{\lambda}$ is a constant very near 1¡£

As a thermal based signal detector, energy resolution of TES is almost constant $FWHM_{r}=2.35*\delta_{r}=2.35*\sqrt{4k_{B}T_{0}^{2}C/\alpha_{I}\sqrt{n/2}}$ if TES work in linear dynamic range \cite{astronomyapply8}. That is, $\delta_{r}^{2}$ is not changed with $\overline{n_{t}}$.

So£¬$a=\delta_{\lambda}^{2}$ and $b=\delta_{r}^{2}$ in Eq (\ref{eq2}) is constant£¬and we can simplify Eq (\ref{eq2}):
\begin{equation}
\delta_{n_t}^{2}=(FWHM_{\overline{n_t}}/2.35)^{2}=a*\overline{n_t}+b \label{eq3}
\end{equation}

According to Eq (\ref{eq3})£¬we can measure $\delta_{n_t}^{2}$ at different $\overline{n_t}$£¬linear fitting it£¬square root the intercept and multiply 2.35, then we get the energy resolution of TES.

\section{energy resolution measurement}
\subsection{TES test}
The TES resistance as a function of temperature, called RT curve is shown in Fig \ref{fig2}. From Fig \ref{fig2} we can get width of RT curve is about 3mK£¬transition temperature is about 542mK.
\begin{center}
\includegraphics[width=8cm]{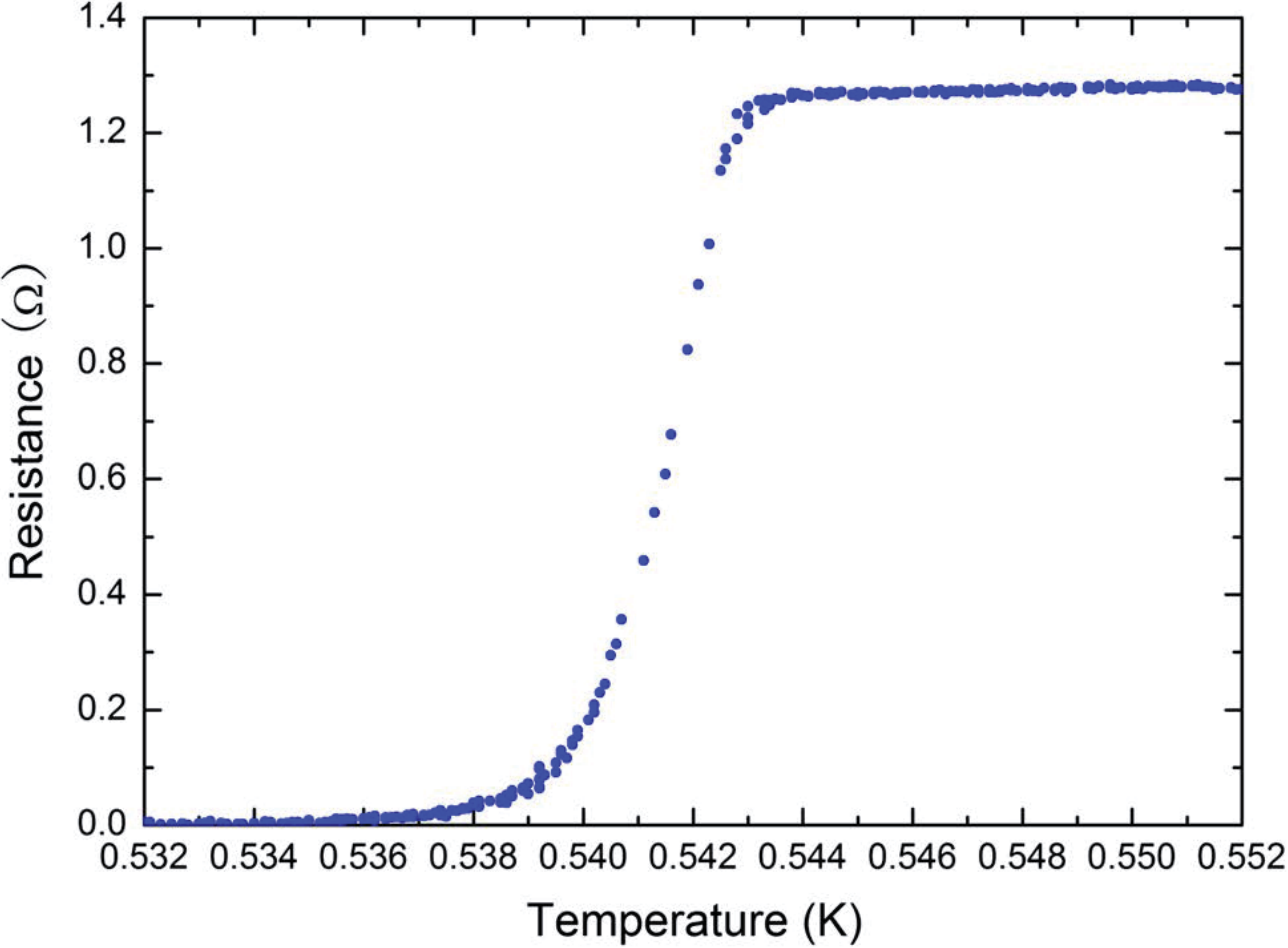}
\figcaption{\label{fig2}   Ti/Al TES RT curve.}
\end{center}
\subsection{pulse hight vs $\overline{n_t}$ }
We measured pulse hightes at different $\overline{n_t}$£¬and they show excellent linearity£¬that's mean TES is working at a linear work point of. By pulse hight and shape, we can get the TES current and resistance. Then we can calculate the energy of each pulse by integral the power change and time.

\begin{center}
\includegraphics[width=8cm]{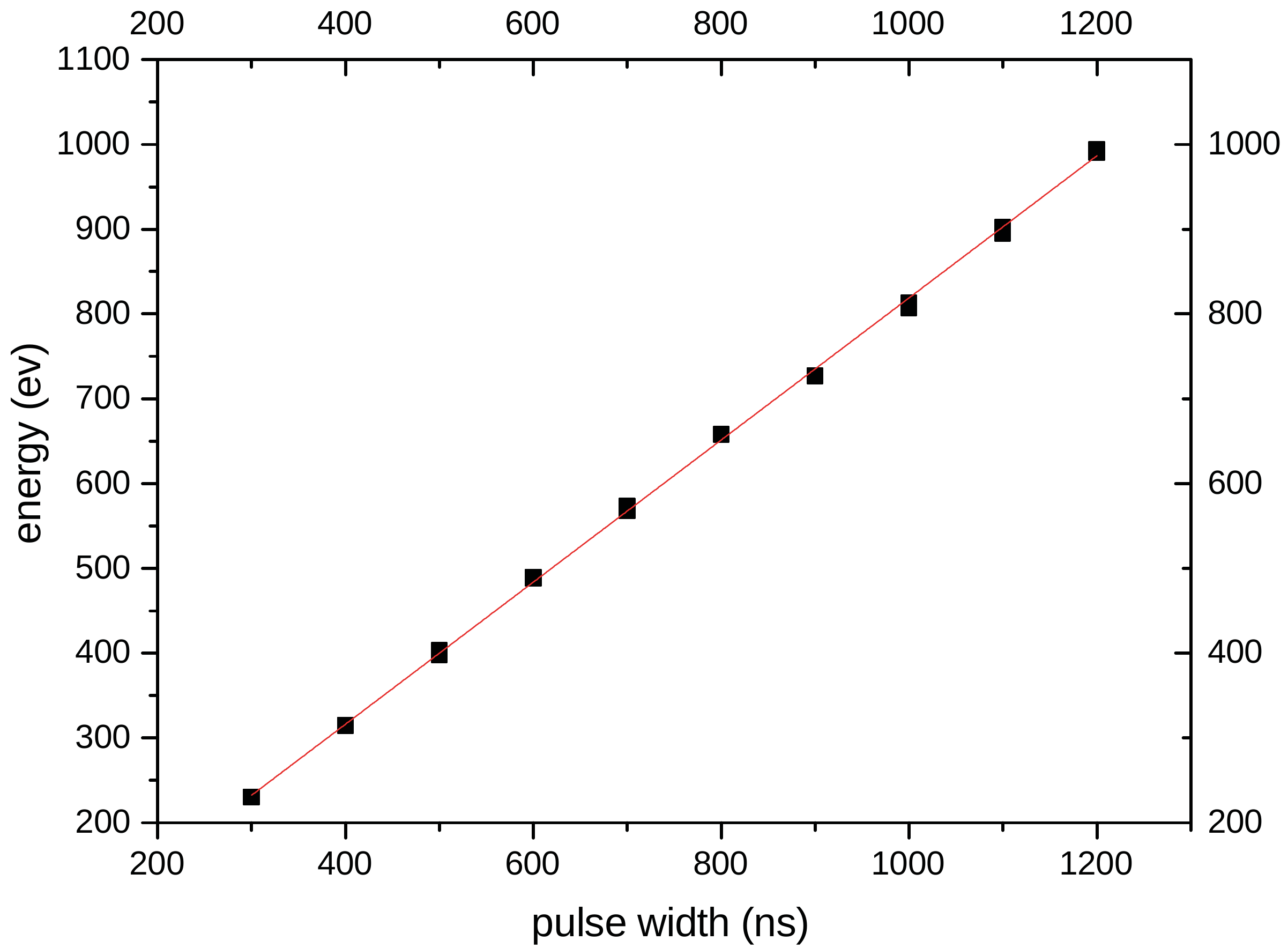}
\figcaption{\label{fig3} pulse hight at different $\overline{n_t}$}
\end{center}
\subsection{energy spectrum under different input energy}
We measured energy spectrum under different $\overline{n_t}$£¬we can see the $FWHM_{\overline{n_t}}$ of the spectrum increase as $\overline{n_t}$ increasing, agree with the theory prediction.
\begin{center}
\includegraphics[width=8cm]{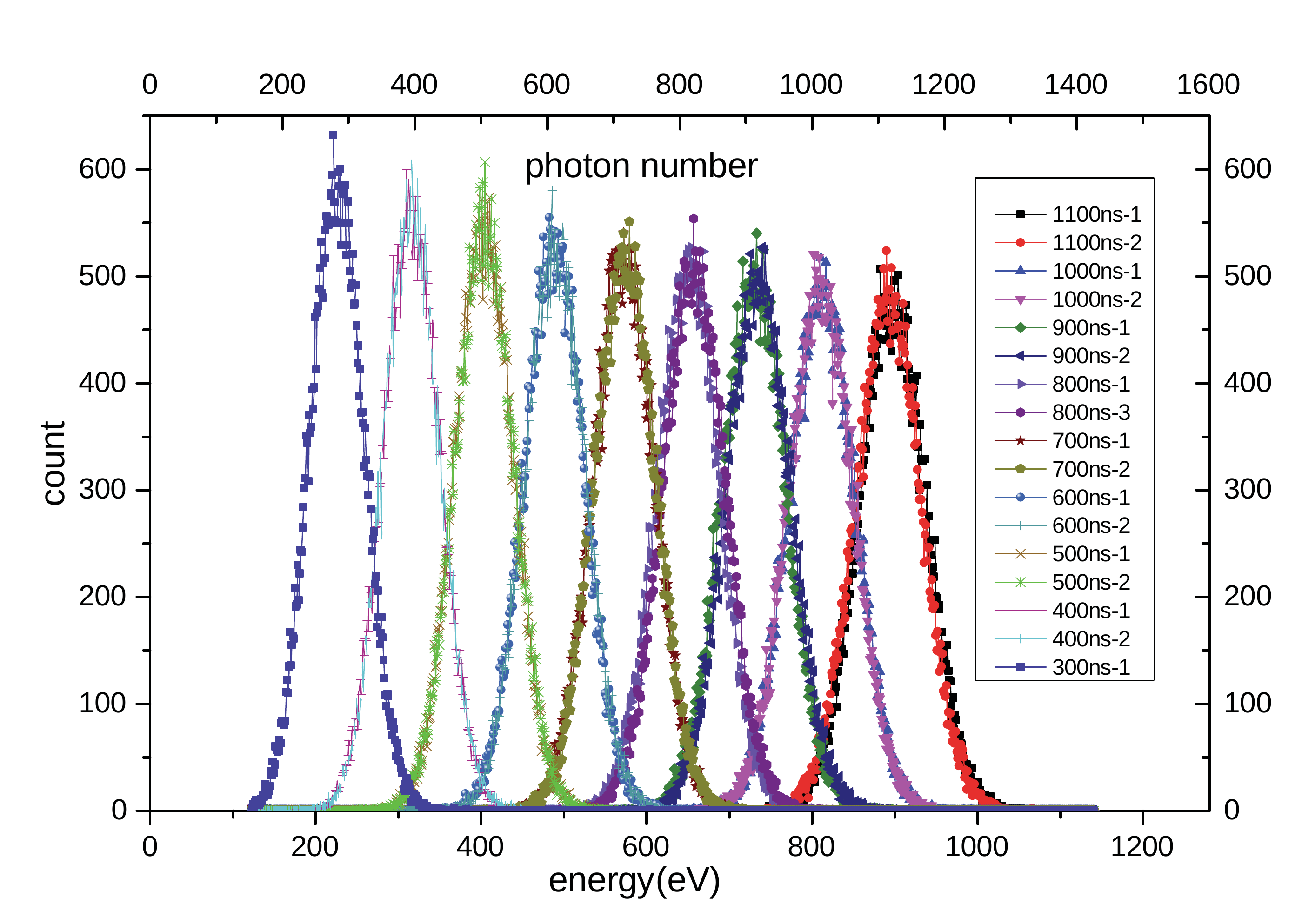}
\figcaption{\label{fig4}   energy spectrum under different $\overline{n_t}$}
\end{center}
\section{Result and Discussion}
\subsection{square of energy resolution vs $\overline{n_t}$}
By analysing the energy spectrum£¬we can get ${FWHM_{\overline{n_t}}}^2$ under different $\overline{n_t}$, the result is shown in Fig \ref{fig5},  Eq (\ref{eq3}) agree with the result very well. We get $\delta_{\lambda}^{2}=0.992\pm0.022$ agree with the theory prediction $\delta_{\lambda}^{2}=1$. $\delta_r=38.295\pm0.488$ corresponding to a  $\delta{E}=72.30\pm0.39eV$ energy resolution.
\begin{center}
\includegraphics[width=8cm]{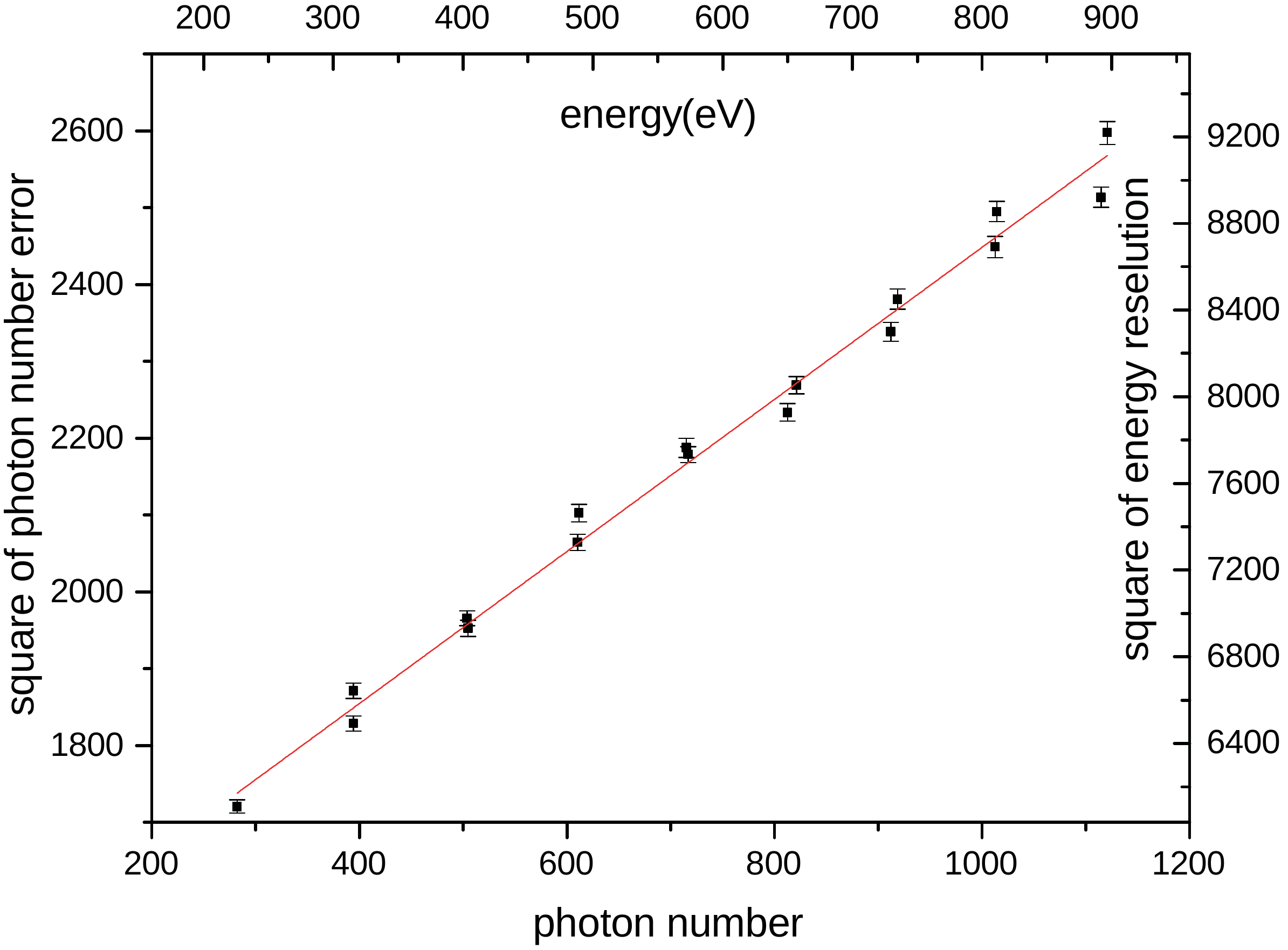}
\figcaption{\label{fig5}  $\delta{{\overline{n_t}}^2}$ under different $\overline{n_t}$}
\end{center}

\subsection{future work}
The experiment results agree very well with the theory predictions, but 72eV is not a very good energy resolution. The main reason is as the following description:

First energy resolution is affected wiby temperature, $FWHM_{r}=2.35*\delta_{r}=2.35*\sqrt{4k_{B}T_{0}^{2}C/\alpha_{I}\sqrt{n/2}}$. Our He3 refrigerator can only work on temperature higher than 400mK. If we can reach lower temperature, we may get better resolution.

Secondly, we use Multichannel Signal Analyzer (MSA) to get the energy spectrum directly, this is not the best method to get the energy resolution, so in the future, we need design a new data acquisition system to improve the energy resolution.

\section{acknowledgment.}
Fangjun Lu give a lot of instruction in the statistical method, We acknowledge the support of CAS Strategic Priority A(Grant No. XDA040762), State Key Development Program for Basic Research of China(Grant No.2011CBA00304) and Tsinghua University Initiative Scientific Research Program(Grant No.20131089314)

\vspace{-1mm}
\centerline{\rule{80mm}{0.1pt}}

\end{multicols}{2}
\clearpage
\end{document}